\documentclass[p3,12pt,number]{elsarticle}
 
 \usepackage{graphicx}
 \usepackage{amssymb}
\newcommand{\vect}[1]{\mathbf{#1}}

\begin{document}
\begin{frontmatter}
\title{Algorithm for calculating spectral intensity due to charged particles in arbitrary motion}
\author{A.G.R. Thomas}
\address{Centre for Ultrafast Optical Science, University of Michigan, Ann Arbor, Michigan, US.}
\ead{agrt@umich.edu}
\date{\today}

%===============================================================================0
%               ABSTRACT
%===============================================================================0
\begin{abstract}
An algorithm for calculating the spectral intensity of radiation due to the coherent addition of many particles with arbitrary trajectories is described. Direct numerical integration of the Li\'enard-Wiechert potentials,  in the far-field, for extremely high photon energies and many particles is made computationally feasible by a mixed analytic and numerical method. Exact integrals of spectral intensity are made between discretely sampled trajectories, by assuming the space-time four-vector is a quadratic function of proper time. The integral Fourier transform of the trajectory with respect to time, the modulus squared of which comprises the spectral intensity, can then be formed by piecewise summation of exact integrals between discrete points. Because of this, the calculation is not restricted by discrete sampling bandwidth theory, and hence for smooth trajectories, time-steps many orders larger than the inverse of the frequency of interest can be taken.
\end{abstract}

\end{frontmatter}

%===============================================================================0
%               INTRODUCTION
%===============================================================================0
%
\section{Introduction}
Radiation from synchrotrons is a well developed field, and a number of numerical methods for calculating the radiation spectrum exist \cite{ISI:000255096304159, ISI:000244647900082, ISI:000180122000017,ISI:000177920400002,ISI:000087317900067}. Laser driven sources of radiation have also triggered interest in measurements \cite{ISI:000224131400049, ISI:000234608300041,ISI:000235746900015,ISI:000234608300040,ISI:000242538700029, ISI:000244646100041,ISI:000256206400035,ISI:000254024500038, ISI:000253724500017} and numerical calculations \cite{  ISI:000224131400048,ISI:000258927900002, ISI:000232109000057,ISI:000259615700003} of radiation, particularly in the area of laser-plasma particle acceleration. As sources of high-energy particle beams and radiation, laser-plasma based techniques may be used for a large range of future applications. Ultrafast X-ray sources would be useful in, for example, time resolved diffraction, medical imaging, spectroscopy and microscopy of transient physical, chemical, or biological phenomena. Laser wakefield acceleration \cite{ISI:A1979HD13400009} of high energy electrons beams has recently successfully demonstrated the production of GeV peak energy electron beams \cite{ISI:000241493100020}, and has become a highly cited field of research. Such beams could be used directly for radiotherapy or radiographic imaging, or alternatively can be converted into fs duration, high brightness sources of x-rays. Numerical calculation of the x-ray spectrum is therefore of interest. Scattering of laser pulses from relativistic electron beams is another area which may require well characterized angularly resolved spectra from a realistic bunch, for comparison with experiment.

In general, the trajectories of particles in laser driven experiments are quite complicated; wakefields, electron beam interactions with intense laser fields, electron orbits in laser generated channels and in laser-solid interactions all represent sources of radiation. The radiation fields can be explicitly calculated by a fast Fourier transform method or by finite differencing of the Li\'enard-Wiechert fields in time, but these are computationally intensive processes due to the constraints of the Whittaker-Shannon-Nyquist sampling theorem \cite{ ISI:000174400500008,ISI:A1949UM84500004}. This essentially states that for discretely sampled data, frequencies higher than half the sampling frequency are aliased to frequencies lower than half the sampling frequency. Since the spectral power as a function of frequency emitted is essentially equivalent to a Fourier transform, the highest resolvable frequency in the spectrum is constrained. However, for relativistic particles radiation can be produced at much higher frequencies than the actual time-scale corresponding to the change in momentum of the particles, because the radiation co-propagates with the particle.

Here, an algorithm is developed which uses a combination of numerical and analytic methods to integrate the Li\'enard-Wiechert fields, for arbitrary trajectories of many charged particles, to frequencies greatly exceeding the Nyquist frequency, $\nu_N$. This means that the sampling rate to accurately reproduce a particular spectrum can be many orders of magnitude lower than $1/\nu_N$, and therefore faster to solve numerically. This is particularly relevant to calculations of radiation from betatron oscillations in laser wakefield accelerators and beam-laser interactions, but the technique is applicable to numerous other areas of physics.
%===============================================================================0
%               THE MODEL
%===============================================================================0
\section{The numerical method}
The spectral intensity of radiation emitted by a number $N_P$ of accelerating point charges, with the $j$th particle at position $\vect{r}_j$ and with normalized velocity $\vect{\beta}_j=\vect{v}_j/c$, can be expressed, in the far-field, as:
 \begin{equation}
\frac{d^2I}{d\omega d\Omega}=\frac{\mu_0e^2c}{16\pi^3}\omega^2\Bigg|{\int_{-\infty}^{\infty}\sum_{j=1}^{N_P}
\vect{\hat{s}}\times\vect{\beta}_je^{i\omega (t -\vect{\hat{s}}\cdot \vect{r}_j/c)}}dt\Bigg|^2\;,
\label{spec:Jackson} 
\end{equation}
where the unit vector $\vect{\hat{s}}$ is in the direction of observation, at a distance far compared with the scale of the emission region. This can be written alternatively in terms of proper time, $\tau$:
 \begin{equation}
\frac{d^2I}{d\omega d\Omega}=\frac{\mu_0e^2c}{16\pi^3}\omega^2\Bigg|\sum_{j=1}^{N_P}{\int_{-\infty}^{\infty}\vect{\hat{s}}\times\vect{v}_je^{i\kappa_\alpha x_j^\alpha}}d\tau\Bigg|^2\;,
\label{spec:Jackson_proper} 
\end{equation}
where the four-wave-vector $\kappa_\alpha=\omega \{ 1,\vect{\hat{s}}/c\}$, and $\vect{v}_j$ is the momentum part of the $j$th particle's four-velocity defined as:
\begin{equation}
v_j^\alpha=\frac{dx_j^\alpha}{d\tau}\;.
\end{equation}
One of the advantages of using proper time rather than `laboratory' time for numerical calculations is that for a uniform step finite differencing scheme, the time resolution is effectively adaptive; as the particle gains inertia and is therefore accelerated at a decreased rate for a similar force, the {\it laboratory} time step-size increases. To numerically integrate the equations of motions for charged particles, both $x^\alpha$ and $\vect{v}$ have to be recorded at a number of discrete points. To then perform the spectral integration numerically, a `zeroth order' model would be to reduce equation \ref{spec:Jackson_proper} to a summation over finite differenced points:
 \begin{equation}
\frac{d^2I}{d\omega d\Omega}=\frac{\mu_0e^2c}{16\pi^3}\omega^2\Bigg|{\sum_{j=1}^{N_P}\sum_{n=0}^{N_\tau}\vect{\hat{s}}\times \vect{v}_j^ne^{i\kappa_\alpha x_j^{\alpha,n}}}\Delta\tau\Bigg|^2\;.
\label{spec:Jackson_proper_diff} 
\end{equation}
The problem with this method is that frequencies higher than half the sampling rate are aliased to lower frequencies, and therefore an upper limit is put on the maximum frequency that can be effectively resolved to $\sim1/\Delta \tau$ \cite{ISI:000174400500008,ISI:A1949UM84500004}. For attempts to simulate high energy photons from laser interactions, this can be computationally prohibitive. However, the motion of the particles which lead to such high-energy photons generally consists of changes on timescales much larger than the radiation frequencies produced. Hence, a different approach is taken, which is to assume that the motion of a particle between time-steps can be approximated by an interpolating function. The spectral integral can be expressed as a summation over analytic integrals between each time-step:
 \begin{equation}
\frac{d^2I}{d\omega d\Omega}=\frac{\mu_0e^2c}{16\pi^3}\omega^2\Bigg|{\sum_{j=1}^{N_P}\sum_{n=0}^{N_\tau}\vect{\hat{s}}\times
 \int_{\tau_n-\Delta\tau/2}^{\tau_n+\Delta\tau/2}\vect{v}_j(\tau)e^{i\kappa_\alpha x_j^{\alpha}(\tau)}}d\tau\Bigg|^2
 \;.
\label{spec:Jackson_proper_diff_proper} 
\end{equation}
Here the notation used is that Greek character sub/super-scripts represents components of 4-vector quantities, subscript Roman characters denote either particle number, $j$, or (proper) time-step, $n$, and bold font represents 3-vector quantities. Proper time is discretized into steps of size $\Delta\tau$. An exact analytic solution can then be employed to calculate the sub-integral between time-steps, using  an interpolating function for $\vect{v}_j$ and $x_j^{\alpha}$ between discrete time steps. A quadratic interpolation for $x^\alpha$ can accurately model both linear and harmonic accelerations, and a linear interpolation for the velocity four-vector is consistent. Thus, the four-velocity and four-displacement at a time $\tau-\tau_n$ are approximated by:
\begin{eqnarray}
x^\alpha(\tau)&=&x^\alpha_{0n}+x^{\alpha}_{1n}{(\tau-\tau_n)}+x^\alpha_{2n}{(\tau-\tau_n)}^2\;,
\nonumber\\
\vect{v}_j(\tau)&=&\vect{v}_{0j,n}+\vect{v}_{1j,n}{(\tau-\tau_n)}\;,
\end{eqnarray}
where $x^\alpha_{0n}=x^\alpha(\tau_n)$, $\vect{v}_{0j,n}=\vect{v}_j(\tau_n)$, and $\vect{v}_{1j,n}$, $x^{\alpha}_{1n}$ and $x^{\alpha}_{2n}$ are interpolation coefficients.
Each integral over time in equation \ref{spec:Jackson_proper_diff_proper} can then be broken up into a series of {\it analytic integrals} between time steps from $\tau=n\Delta\tau$ to $\tau=(n+1)\Delta\tau$:
 \begin{equation}
\frac{d^2I}{d\omega d\Omega}=\frac{\mu_0e^2c}{16\pi^3}\omega^2\Bigg|\sum_{j=1}^{N_P}\sum_{n=0}^{N_\tau}\vect{\hat{s}}\times
\tilde{\mathcal{I}}_{j,n}(\omega)\Bigg|^2
 \;,
\end{equation}
where, with quadratic interpolation:
 \begin{eqnarray}
\tilde{\mathcal{I}}_{j,n}(\pm\omega)=\int_{\tau_n-\Delta \tau/2}^{\tau_n+\Delta \tau/2}
\left(\vect{v}_{0j,n}+\vect{v}_{1j,n}{[\tau-\tau_n]}\right)
e^{\pm i\kappa_\alpha (x^\alpha_{0j,n}+x^{\alpha}_{1j,n}{[\tau-\tau_n]}+x^\alpha_{2j,n}{[\tau-\tau_n]}^2)}
d\tau\;,
\end{eqnarray}
and the space and velocity fourvectors are assumed real so that $\tilde{\mathcal{I}}_{j,n}^{\ast}(\omega)=\tilde{\mathcal{I}}_{j,n}(-\omega)$. A change of variables in the integral leads to:
 \begin{eqnarray}
\tilde{\mathcal{I}}_{j,n}(\pm\omega)&=& e^{\pm i\kappa_\alpha x^\alpha_{0j,n}}
\int_{-\Delta \tau/2}^{\Delta \tau/2}
\left(\vect{v}_{0j,n}+\vect{v}_{1j,n}{\tau}\right)
e^{\pm i\kappa_\alpha x^{\alpha}_{1j,n}{\tau}}
e^{\pm i\kappa_\alpha x^\alpha_{2j,n}{\tau}^2}
d\tau\;.
\end{eqnarray}
If each particle trajectory is accurately described by the interpolation function between grid-points, then the total integral is exactly solved. This is the key difference which allows calculation of the spectrum to far beyond the Nyquist frequency corresponding to the time step between known position and moment values. The integral with respect to $\tau$, ${\mathcal{I}}_{j,n}(\omega)=\exp(-{ i\kappa_\alpha x^\alpha_{0j,n}})\tilde{\mathcal{I}}_{j,n}(\omega)$, can be split into real and imaginary parts:
 \begin{eqnarray}
\Re\left({\mathcal{I}}_{j,n}\right)&=&
\int_{-\Delta \tau/2}^{\Delta \tau/2}
\left(\vect{v}_{0j,n}+\vect{v}_{1j,n}{\tau}\right)
\cos\left(\kappa_\alpha x^{\alpha}_{1j,n}{\tau}+\kappa_\alpha x^\alpha_{2j,n}{\tau}^2\right)
d\tau\;,\\
\Im\left({\mathcal{I}}_{j,n}\right)&=&
\pm\int_{-\Delta \tau/2}^{\Delta \tau/2}
\left(\vect{v}_{0j,n}+\vect{v}_{1j,n}{\tau}\right)
\sin\left(\kappa_\alpha x^{\alpha}_{1j,n}{\tau}+\kappa_\alpha x^\alpha_{2j,n}{\tau}^2\right)
d\tau\;.
\end{eqnarray}
It can be shown that these integrals have the solutions:
 \begin{eqnarray}
\Re\left({\mathcal{I}}_{j,n}\right)&=&
\frac{1}{4{\chi_{2j,n}}}
\left\{
\begin{array}{l l}
&\vect{\Psi}_+\left[C(\Theta_+)-C(\Theta_-)\right]\\
+&\vect{\Psi}_-\left[S(\Theta_+)-S(\Theta_-)\right]\\
+&2\vect{v}_{1j,n}\left[\sin\Phi_+-\sin\Phi_-\right]
\end{array}
\right\}
\;,
\label{fresnel1}
\\
\Im\left({\mathcal{I}}_{j,n}\right)&=&
\pm\frac{1}{4{\chi_{2j,n}}}
\left\{
\begin{array}{l l}
&\vect{\Psi}_+\left[S(\Theta_+)-S(\Theta_-)\right]\\
-&\vect{\Psi}_-\left[C(\Theta_+)-C(\Theta_-)\right]\\
-&2\vect{v}_{1j,n}\left[\cos\Phi_+-\cos\Phi_-\right]
\end{array}
\right\}
\;,
\label{fresnel2}
\end{eqnarray}
where $C(x)=\int_0^x\cos(\pi t^2/2)dt$ and $S(x)=\int_0^x\sin(\pi t^2/2)dt$ are the Fresnel integrals, and:
\begin{eqnarray}
\Theta_\pm &=& \frac{\chi_{1j,n}\pm\chi_{2j,n}\Delta\tau}{\sqrt{2\pi\chi_{2j,n}}}\;,\\
\vect{\Psi}_+ &=& \sqrt{\frac{2\pi}{\chi_{2j,n}}}\left(2{\chi_{2j,n}} \vect{v}_{0j,n}-\chi_{1j,n}\vect{v}_{1j,n}\right)
\cos\left(\frac{{\chi_{1j,n}}^2}{4\chi_{2j,n}}\right)\;,\\
\vect{\Psi}_- &=& \sqrt{\frac{2\pi}{\chi_{2j,n}}}\left(2{\chi_{2j,n}} \vect{v}_{0j,n}-\chi_{1j,n}\vect{v}_{1j,n}\right)
\sin\left(\frac{{\chi_{1j,n}}^2}{4\chi_{2j,n}}\right)\;,\\
\Phi_\pm &=& \frac{\Delta\tau^2}{4}\chi_{2j,n}\pm\frac{\Delta\tau}{2}\chi_{1j,n}\;,\\
\chi_{1j,n} &=& \kappa_\alpha x^{\alpha}_{1j,n}\;,\\
\chi_{2j,n} &=& \kappa_\alpha x^{\alpha}_{2j,n}\;.
\end{eqnarray}
Here, the Fresnel integrals are solved numerically by using the power series and continued fraction expressions in {\it Numerical Recipes in C++} \cite{citeulike:1321549}. Although the benefit gained in using less time steps by this method far outweighs the cost of calculating these functions, this is a computationally expensive process, and if the use of Fresnel integrals can be avoided it would be beneficial. If the second exponent in the integral is small, it is appropriate to Taylor expand the exponential function and truncate at order $\tau^2$, hence:
 \begin{eqnarray}
{\mathcal{I}}_{j,n}(\pm\omega)&=& %e^{\pm i\kappa_\alpha x^\alpha_{0j,n}}
\int_{-\Delta \tau/2}^{\Delta \tau/2}
\left(\vect{v}_{0j,n}+\vect{v}_{1j,n}{\tau}\right)
e^{\pm i\kappa_\alpha x^{\alpha}_{1j,n}{\tau}}
\left(1\pm i\kappa_\alpha x^\alpha_{2j,n}{\tau}^2+\dots\right)
d\tau\;.
\nonumber\\
&\simeq&%e^{\pm i\kappa_\alpha x^\alpha_{0j,n}}
\int_{-\Delta \tau/2}^{\Delta \tau/2}
\left(\vect{v}_{0j,n}+\vect{v}_{1j,n}{\tau}\pm i\chi_{2j,n}\vect{v}_{0j,n}{\tau}^2\right)
e^{\pm i\chi_{1j,n}{\tau}}
d\tau\;.
\end{eqnarray}
The integrals when $x^\alpha_{2j,n}{\tau}^2$ is small are:
 \begin{eqnarray}
\Re\left({\mathcal{I}}_{j,n}\right)&=&
\vect{v}_{0j,n}
{\mathcal{I}}_{0j,n}
\;,
\label{realfirstorder}
\\
\Im\left({\mathcal{I}}_{j,n}\right)&=&
\vect{v}_{1j,n}
{\mathcal{I}}_{1j,n}
\pm \chi_{2j,n}\vect{v}_{0j,n}
{\mathcal{I}}_{2j,n}
\;,
\end{eqnarray}
where:
\begin{eqnarray}
{\mathcal{I}}_{0j,n}&=&{\rm sinc}\left(\frac{\chi_{1j,n}\Delta\tau}{2}\right)\Delta\tau\;,
\\
{\mathcal{I}}_{1j,n}&=& \frac{\Delta\tau}{\chi_{1j,n}}\left[
{\rm sinc}\left(\frac{\chi_{1j,n}\Delta\tau}{2}\right)
- \cos\left(\frac{\chi_{1j,n}\Delta\tau}{2}\right)
\right]\;,\\
{\mathcal{I}}_{2j,n}&=&%\frac{2\Delta\tau}{(\kappa_\alpha x^{\alpha}_{1j,n})^2}
%\left[\cos\left(\kappa_\alpha x^{\alpha}_{1j,n}\frac{\Delta\tau}{2}\right)
%-{\rm sinc}\left(\kappa_\alpha x^{\alpha}_{1j,n}\frac{\Delta\tau}{2}\right)\right]
\frac{\Delta\tau^3}{4}{\rm sinc}\left(\frac{\chi_{1j,n}\Delta\tau}{2}\right)
-\frac{2}{\chi_{1j,n}}{\mathcal{I}}_{1j,n}
\;,
\end{eqnarray}
where $ {\rm sinc}(x)$ is the unnormalized cardinal sine function, $ {\rm sinc}(x)=\sin(x)/x$. The Taylor expanded solution for the integral ${\mathcal{I}}_{j,n}$ is significantly faster to solve than the Fresnel integral expressions, so a condition statement can switch between solutions depending on the size of $\chi_{2j,n}{\Delta\tau}^2$. For practical purposes, the expansion solution is necessary because the exact solution, when numerically solved, diverges for small $\chi_{2j,n}{\Delta\tau}^2$ due to floating point truncation error, the accuracy of the trigonometric functions and the inverse $\chi_{2j,n}{\Delta\tau}^2$ relationship. Some threshold value, $T$, for $\chi_{2j,n}{\Delta\tau}^2$ can be used to choose between the two models. This must be between 1 and 0, as the Taylor expansion is only valid for $\chi_{2j,n}{\Delta\tau}^2<1$. For values less than the threshold, $\chi_{2j,n}{\Delta\tau}^2<T$, the Taylor expansion solution is used and for values greater than threshold, $\chi_{2j,n}{\Delta\tau}^2>T$, the exact solution is used. In addition, {\it long double} precision floating point numbers are necessary for sufficiently accurate calculation of the functions. A good threshold value is $T=10^{-3}$, as a balance between the accuracy of the solution and the speed of the algorithm.

\begin{figure}[htbp]
\begin{center}
\includegraphics[width=5in]{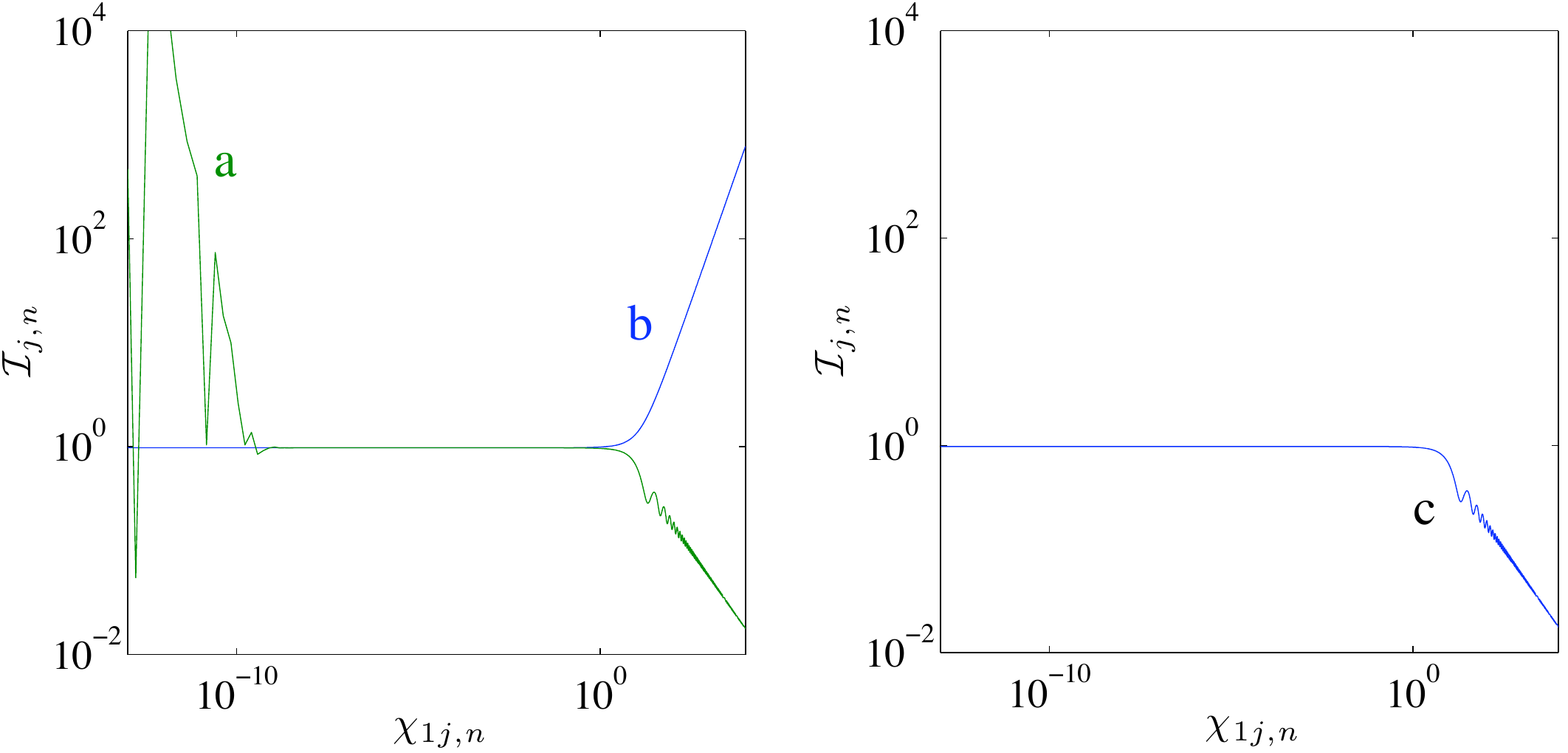}
\caption{Various model calculations of integral $\mathcal{I}_{j,n}$ for $d\tau=1$, $v_{0j,n}=1$, $v_{1j,n}=2$, $\chi_{1j,n}=1$ and varying $\chi_{2j,n}$. (a) Calculation of integral using fresnel integral form. (b) Calculation using Taylor series expansion form of integral. (c) Calculation using finite differenced integral with $10^7$ steps and Simpson's rule.}
\label{figure1}
\end{center}
\end{figure}

Figure \ref{figure1} shows calculations of integral $\mathcal{I}_{j,n}$ for $d\tau=1$, $v_{0j,n}=1$, $v_{1j,n}=2$, $\chi_{1j,n}=1$ and varying $\chi_{2j,n}$. (a) Calculation of integral using Fresnel integral form. (b) Calculation using Taylor series expansion form of integral. (c) Calculation using finite differenced integral with $10^7$ steps and Simpson's rule. It is clear in this specific example that the Fresnel integral solution is correct for all but the smallest values of $\chi_{2j,n}$. For these values, since the equations \ref{fresnel1} and \ref{fresnel2} involve the subtraction of large terms and division by $\chi_{2j,n}$, the truncation errors are amplified to significant values. The Taylor expanded solution does not contain a term $1/\chi_{2j,n}$, and therefore approaches the correct result for $\chi_{2j,n}\ll 1$.

The radiated spectral intensity is then given by:
 \begin{eqnarray}
\frac{d^2I}{d\omega d\Omega}=\frac{\mu_0e^2c}{16\pi^3}\omega^2\left\{
\begin{array}{l l}
\left[\sum_{j=1}^{N_P}\sum_{n=0}^{N_\tau}\vect{\hat{s}}\times\left(
\Re\left({\mathcal{I}}_{j,n}\right)\cos({\kappa_\alpha x^\alpha_{0j,n}})-\Im\left({\mathcal{I}}_{j,n}\right)\sin({\kappa_\alpha x^\alpha_{0j,n}})\right)\right]^2\\
+\left[\sum_{j=1}^{N_P}\sum_{n=0}^{N_\tau}\vect{\hat{s}}\times\left(
\Im\left({\mathcal{I}}_{j,n}\right)\cos({\kappa_\alpha x^\alpha_{0j,n}})+\Re\left({\mathcal{I}}_{j,n}\right)\sin({\kappa_\alpha x^\alpha_{0j,n}})\right)\right]^2\\
\end{array}
\right\} \;.
\end{eqnarray}
The radiation would normally be considered in a spherical polar coordinate system, $\{r,\theta,\phi\}$, where $\theta$ is the azimuth and $\phi$ is the polar angle. For simplicity, a cartesian coordinate system, $x,y,z$, can be chosen such that the radiation is calculated at an angle $\theta$ with respect to the $z$ axis in the $y-z$ plane. To observe radiation in a particular direction in  $\phi$, the coordinate system can be rotated about the $z$ axis. Defining $x,y,z$ components of combinations of integrals ${\mathcal{I}}_{j,n}$:
 \begin{eqnarray}
 \Re\left(S_{x,y,z}\right)&=&\left\{\sum_{j=1}^{N_P}\sum_{n=0}^{N_\tau}\left(\Re\left({\mathcal{I}}_{j,n}\right)\cos({\kappa_\alpha x^\alpha_{0j,n}})-\Im\left({\mathcal{I}}_{j,n}\right)\sin({\kappa_\alpha x^\alpha_{0j,n}})\right)\right\}_{x,y,z}\;,\\
 \Im\left(S_{x,y,z}\right)&=&\left\{\sum_{j=1}^{N_P}\sum_{n=0}^{N_\tau}\left(\Im\left({\mathcal{I}}_{j,n}\right)\cos({\kappa_\alpha x^\alpha_{0j,n}})+\Re\left({\mathcal{I}}_{j,n}\right)\sin({\kappa_\alpha x^\alpha_{0j,n}})\right)\right\}_{x,y,z}\;,
 \end{eqnarray}
 the radiated spectral intensity is:
\begin{eqnarray}
\frac{d^2I}{d\omega d\Omega}=\frac{\mu_0e^2c}{16\pi^3}\omega^2\left\{
\begin{array}{l l}
 \Re\left(S_x\right)^2+ \Im\left(S_x\right)^2\\
+\left(\Re\left(S_y\right)\cos\theta- \Re\left(S_z\right)\sin\theta\right)^2\\
+\left(\Im\left(S_y\right)\cos\theta- \Im\left(S_z\right)\sin\theta\right)^2
 \end{array}
\right\} \;.
\end{eqnarray}

%===============================================================================0
%               NUMERICAL TESTS
%===============================================================================0
\section{Numerical benchmarks}
\subsection{Synchrotron radiation}
\begin{figure}[htbp]
\begin{center}
\includegraphics[width=4in]{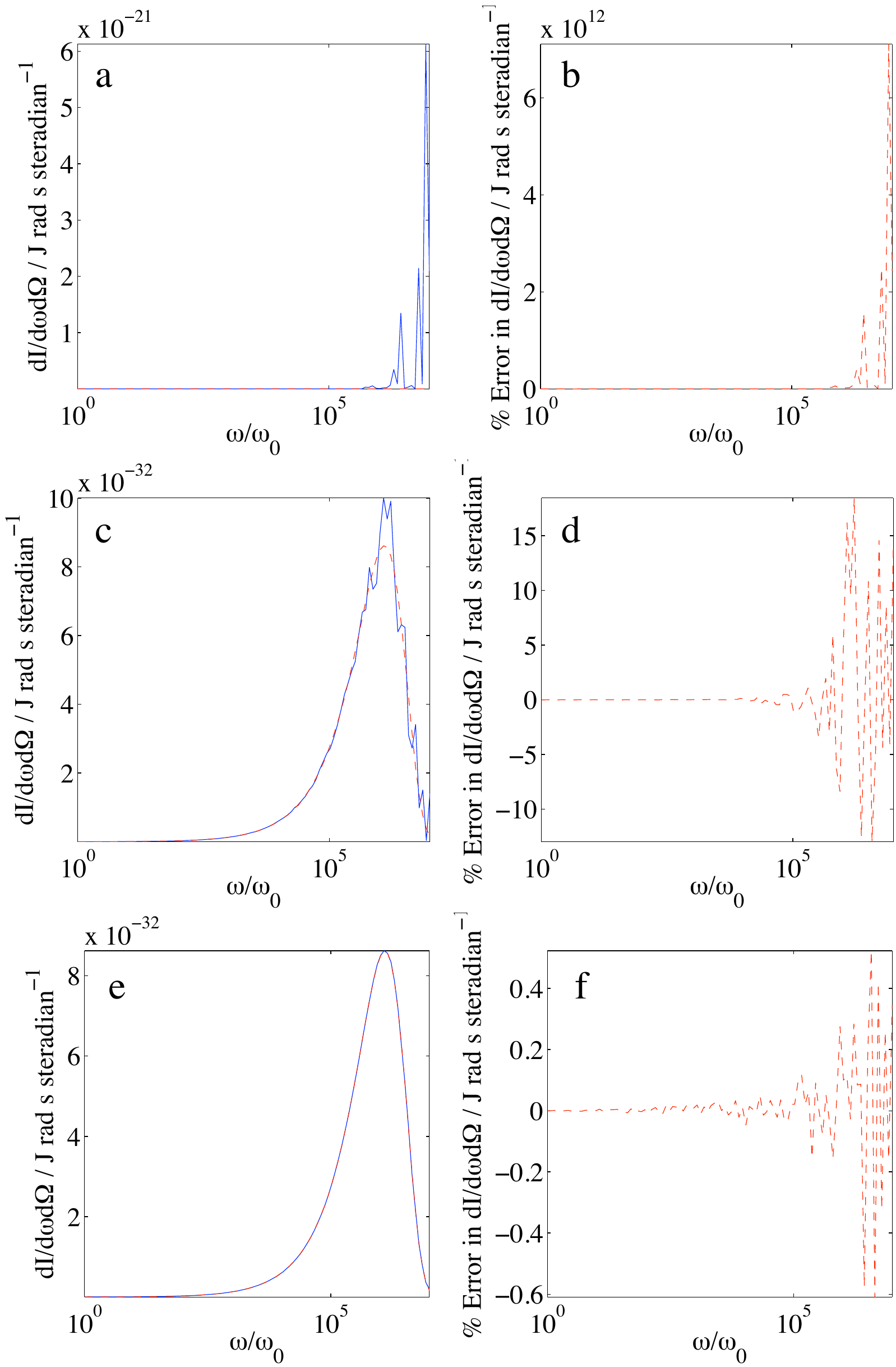}
\caption{Calculated synchrotron spectra (left) and the percentage error relative to the analytic solution (right) for different numerical integration techniques. The electron Lorentz factor is $\gamma=1000$, and the magnetic field strength, $B_0$, is such that $eB_0/m_e\omega_0=1$. The time-step is $\omega_0\Delta\tau=\pi\times10^{-4}$. (a-b) Calculated spectrum and error for finite differenced numerical integration. (c-d) Calculated spectrum and error for summation of exact integrals using first-order interpolation of position. (e-f) Calculated spectrum and error for summation of exact integrals using second-order interpolation of position. In the left-hand images, the red dashed line corresponds to the analytic solution, and the blue solid line corresponds to the numerically calculated solution.}
\label{figure2}
\end{center}
\end{figure}
A test for the algorithm for relativistic motion is reproducing the well known continuum spectrum from a single ultra-relativistic electron rotating in a magnetic field without losing energy. Mathematically this means no radiation damping term in the force equation, physically this can be accelerating electric fields that compensate for the energy losses. An analytic expression for the spectrum is the sum of modified Bessel functions of the second kind, $K_\nu(x)$:
 \begin{eqnarray}
\frac{d^2I}{d\omega d\Omega}&=&\frac{\mu_0e^2c}{12\pi^3} \omega^2\left(\frac{\rho}{c}\right)^2
\left(\frac{1}{\gamma^2}+\theta^2\right)^2
\left[ K_{2/3}^2(\xi)+
 \frac{\theta^2}{\left(\frac{1}{\gamma^2}+\theta^2\right)}
 K_{1/3}^2(\xi)\right]\;,
\end{eqnarray}
where $\xi = {\omega \rho}/3c\gamma^3\left(1+{\theta^2}{\gamma}^2\right)^{3/2}$ and $\rho$ is the radius of curvature. Figure \ref{figure2}  shows calculated synchrotron spectra (left) and the percentage error relative to the analytic solution (right) for different numerical integration techniques. The relative percentage error is defined as the difference between the analytic and numerical solutions divided by the maximum value of the analytic solution.
%\begin{equation}
%E_{\%}=100\frac{\left(\frac{d^2I}{d\omega d\Omega}|_{\rm numerical}-\frac{d^2I}{d\omega d\Omega}|_{\rm analytic}\right)}{\max{\frac{d^2I}{d\omega d\Omega}|_{\rm analytic}}}\;.
%\end{equation}

The electron Lorentz factor is $\gamma=1000$, and the magnetic field strength, $B$, is such that $eB/m_e\omega_0=1$. The time-step is $\omega_0\Delta\tau=\pi\times10^{-4}$, and the maximum frequency calculated is $10^7\omega_0$. Note that this means that the maximum frequency resolved is $\pi\times10^3$ times the inverse of the time-step size. (a-b) show the calculated spectrum and error for finite differenced numerical integration, (c-d) show the spectrum and error for the summation of exact integrals using first-order interpolation of position {\it only} and (e-f) show the spectrum and error for the summation of exact integrals using second-order interpolation of position. In the left-hand images, the red dashed line corresponds to the analytic solution, and the blue solid line corresponds to the numerically calculated solution. It is clear from (a-b) that using the simple finite differencing method (equivalent to a discrete Fourier transform) that the resulting spectrum is of no use whatsoever with this particular time-step. The error grows with frequency to be 12 orders of magnitude larger than the maximum of the spectral intensity that the method is trying to reproduce. Decreasing the time-step size eventually yields an error smaller than 100\%, but to produce accurate spectra, $\omega_0\Delta\tau\ll1$. This is a very limiting factor in calculations of this kind.

By performing a first-order interpolation, as in figure \ref{figure2} (c-d), which corresponds to using the real part of the Taylor expanded solution only, equation \ref{realfirstorder}, the calculated spectrum is clearly now representative of the analytic spectrum, albeit with an error of up to 10\%. The simplicity of this algorithm (the only non algebraic operation involving the calculation of a single sinusoid for each time-step) makes it fast and easy to implement. However, beyond $10^7\omega_0$, the error in the numerical solution starts to grow. Using second-order interpolation, (e-f), yields an error that is at most more than an order of magnitude smaller than the maximum error in the first-order interpolation calculation. Importantly, it is free of spurious oscillations which could lead to misinterpretation of more complex spectra. In addition, the error remains less than the maximum error shown in figure \ref{figure2}(f) up to $5\times10^9\omega_0$, which corresponds to a frequency of  $\omega>10^6/\Delta\tau$. Despite being a more complex algorithm, the second-order method greatly speeds up a spectral calculation due to a larger time-step being allowed. Provided the motion of a particle can be well-described by piecewise quadratic functions (i.e. cubic or higher terms would be small), the method will produce accurate spectra. 
\begin{figure}[htbp]
\begin{center}
\includegraphics[width=5in]{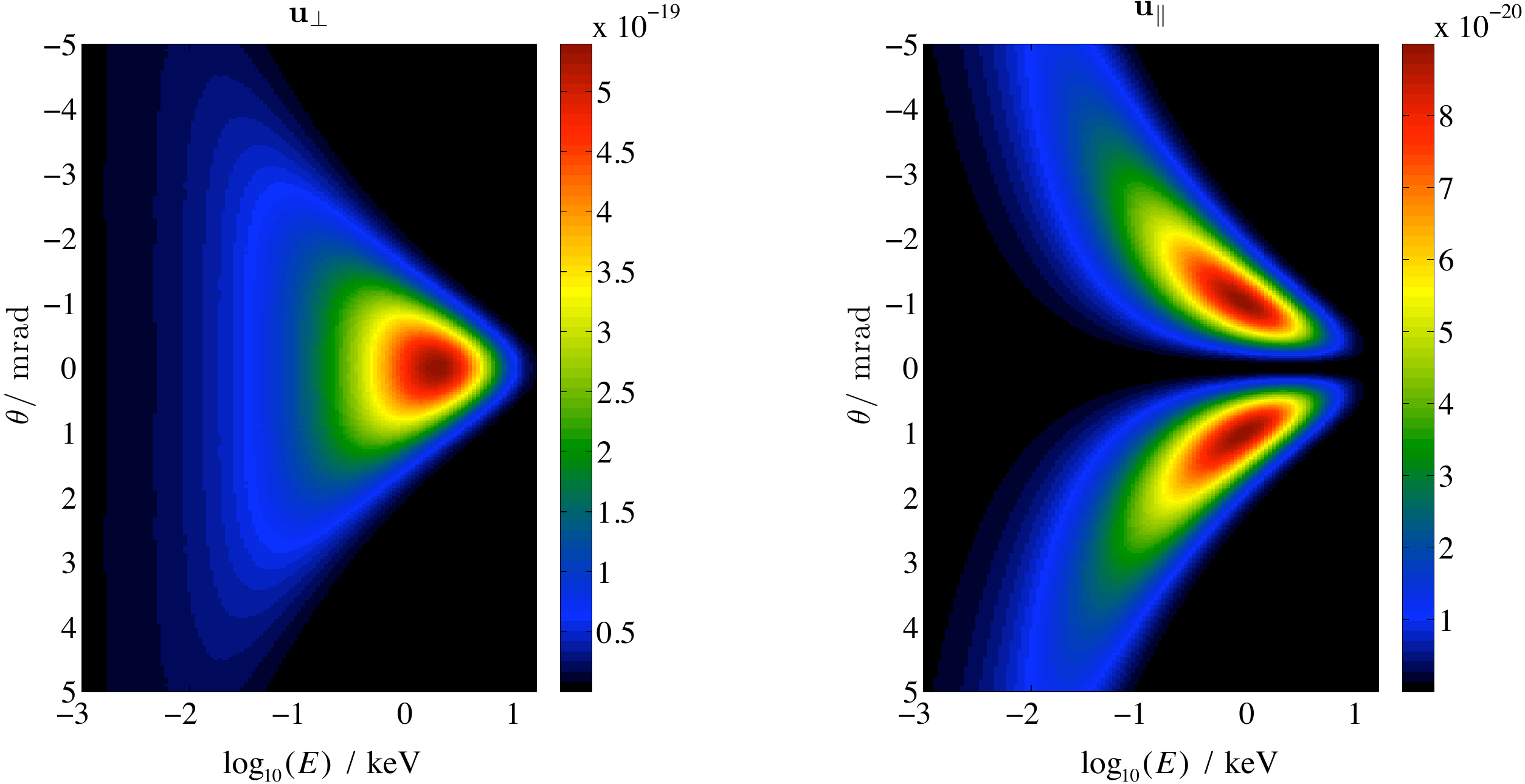}
\caption{Calculated angularly resolved polarization components of synchrotron spectra for an electron Lorentz factor of $\gamma=1000$, and a magnetic field strength of $B_0=13$~T. The time-step is $\omega_0\Delta\tau=\pi\times10^{-4}$. (left) $\vect{u}_\perp$ is the radiated spectral power polarized in the plane of the motion of the electron. (right) $\vect{u}_\parallel$ is the radiated spectral power polarized parallel to the magnetic field.}
\label{figure2d}
\end{center}
\end{figure}

In figure \ref{figure2d}, angularly resolved polarization components of synchrotron spectra for the same parameters are shown. In this case, the frequency scale is shown as an energy scale corresponding to a magnetic field of $B_0=13$~T, which is similar to the parameters of a small synchrotron. The two polarized components of the radiation are $\vect{u}_\perp= {\mu_0e^2c}/{16\pi^3}\omega^2\left(\Re(S_x)^2+ \Im(S_x)^2\right)$ and $\vect{u}_\parallel~=~{\mu_0e^2c}/{16\pi^3}\omega^2\left(\Re\left(S_y\right)\cos\theta- \Re\left(S_z\right)\sin\theta\right)^2
 + \left(\Im\left(S_y\right)\cos\theta- \Im\left(S_z\right)\sin\theta\right)^2$, i.e. the components perpendicular and parallel to the plane of observation. Such calculations can be performed in a matter of minutes on a single processor (the figure shows 500 frequency bins by 500 angular bins resolution and took 30 minutes to calculate on a single processor of a $2\times1.4$~GHz iMac, but a calculation of 100 frequency bins by 100 angular  bins resolution took $\sim$1 minute, including the calculation of the particle trajectory.) 

%%%%%%%%%
\subsection{Linear and non-linear Thomson scattering}
\begin{figure}[htbp]
\begin{center}
\includegraphics[width=3in]{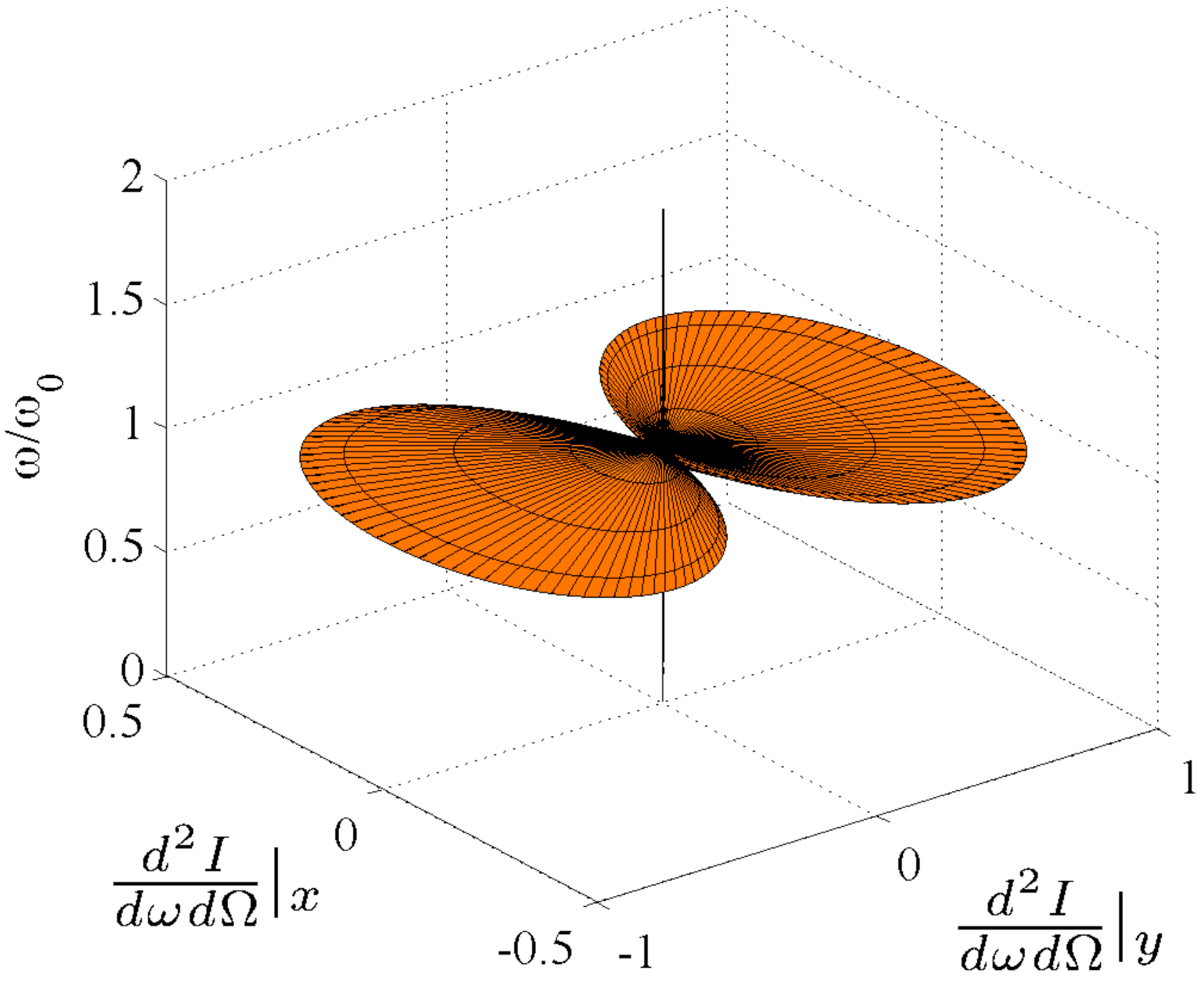}
\caption{Polar plot of radiation emitted by an electron oscillating in a linearly polarized plane electromagnetic wave, with low electric field strength (having a normalized vector potential of $a_0=0.01$) with the field vector along the $x$ axis, calculated numerically. The vertical axis is the frequency normalized to the laser frequency, and the horizontal plane gives the radiated intensity. The angular shape of the structure is a set of $\sin^2\Theta$ lobes, in agreement with the analytic expression. }
\label{figure3}
\end{center}
\end{figure}
To test non-relativistic motion extending to moderately relativistic motion, Thomson scattering of a laser pulse from a single electron can be analyzed. In non-relativistic Thomson scattering from a linearly polarized plane wave, an analytic solution for the radiated power is possible. The field strength corresponding to the interaction is parameterized by the normalized peak vector potential $a_0 = eE_0/mc\omega_0$, where $E_0$ is the peak electric field strength and $\omega_0$ is the angular frequency of the laser. Returning to equation \ref{spec:Jackson} for a single particle, and inserting a time varying velocity $\beta=a_0\sin(\omega_0t)$, and using a coordinate system so that angle $\Theta$ is measured between the polarization axis and the observation direction, the spectral power is:
 \begin{equation}
\frac{d^2I}{d\omega d\Omega}=\frac{\mu_0e^2c}{16\pi^3}\omega^2\Bigg|{\int_{-\infty}^{\infty}
a_0\sin\Theta \sin(\omega_0t)e^{i\omega t -a_0\cos\Theta\cos(\omega_0t)}}dt\Bigg|^2\;.
\end{equation}
To first-order in $a_0$, assuming $a_0\ll1$, this can be integrated to give:
 \begin{equation}
\frac{d^2I}{d\omega d\Omega}=\frac{\mu_0e^2c}{16\pi^2}\omega^2a_0^2\sin^2\Theta
\delta(\omega-\omega_0)\;.
\end{equation}

Here, the Dirac delta function $\delta(x)$ is a representation of $(k/\pi){\rm sinc}^2(x/k)$ in the limit that $k\rightarrow\infty$. The radiated energy is therefore distributed at a single frequency with an angular structure consisting of a $\sin^2\Theta$ shape. In figure \ref{figure3}, the angular distribution of spectral power is shown from calculations using the quadratic algorithm with a time-step of $\omega_0\Delta\tau=\pi/25$, with a plane electromagnetic wave polarized along the $x$-axis with a field strength of $a_0=0.01$. Time is integrated to include 20 wave-periods. The angular shape of the structure is a set of $\sin^2\Theta$ lobes, in agreement with the analytic expression. Although the spectrum as a function of energy is a peaked distribution centered at $\omega=\omega_0$, it is not a Dirac delta function. However, the analytic solution is for an infinite summation of wave-periods, whereas for obvious reasons a finite summation is calculated numerically. In terms of frequency, the calculation yields a ${\rm sinc}^2(\omega-\omega_0)$ shape, characteristic of a finite window, which approaches the Dirac delta as the size of the window is increased. 

As the field strength increases in the interaction, the electron motion becomes more complicated due to the magnetic field, and tends towards a figure of eight motion. This means that additional harmonics of the motion appear in the spectrum. For the case of exact back-scatter from an electron with an initial Lorentz factor of $\gamma_0$, the motion of the electron means that it experiences a Doppler shifted plane wave, and therefore the period of oscillation is modified. In reference \cite{ISI:000182450200081} an analytic expression for the shifted fundamental frequency, $\omega_1$, was given as:
 \begin{equation}
\frac{\omega_1}{\omega_0}=\left(\frac{2}{2+a_0^2}\right)\gamma_0^2\left(1-\beta_0\right)^2
\;,
\end{equation}
where $\beta_0$ is the initial velocity of the particle, and the power per unit solid angle in the $m$th harmonic of $\omega_1$ is:
 \begin{eqnarray}
p_m=
\Bigg\{
\begin{array}{c l}
\frac{\mu_0e^2c}{16\pi}\left(\frac{a_0^2\omega_0^2}{\gamma_0^2\left(1-\beta_0\right)^2}\right)
\frac{\omega_1^4}{\omega_0^4}
\left[J_{(m-1)/2}(m\zeta)-J_{(m+1)/2}(m\zeta)\right]^2& \quad m {\rm~odd}\;,\\
0 & \quad m {\rm~even}\;,
\end{array}
\end{eqnarray}
\begin{figure}[htbp]
\begin{center}
\includegraphics[width=3in]{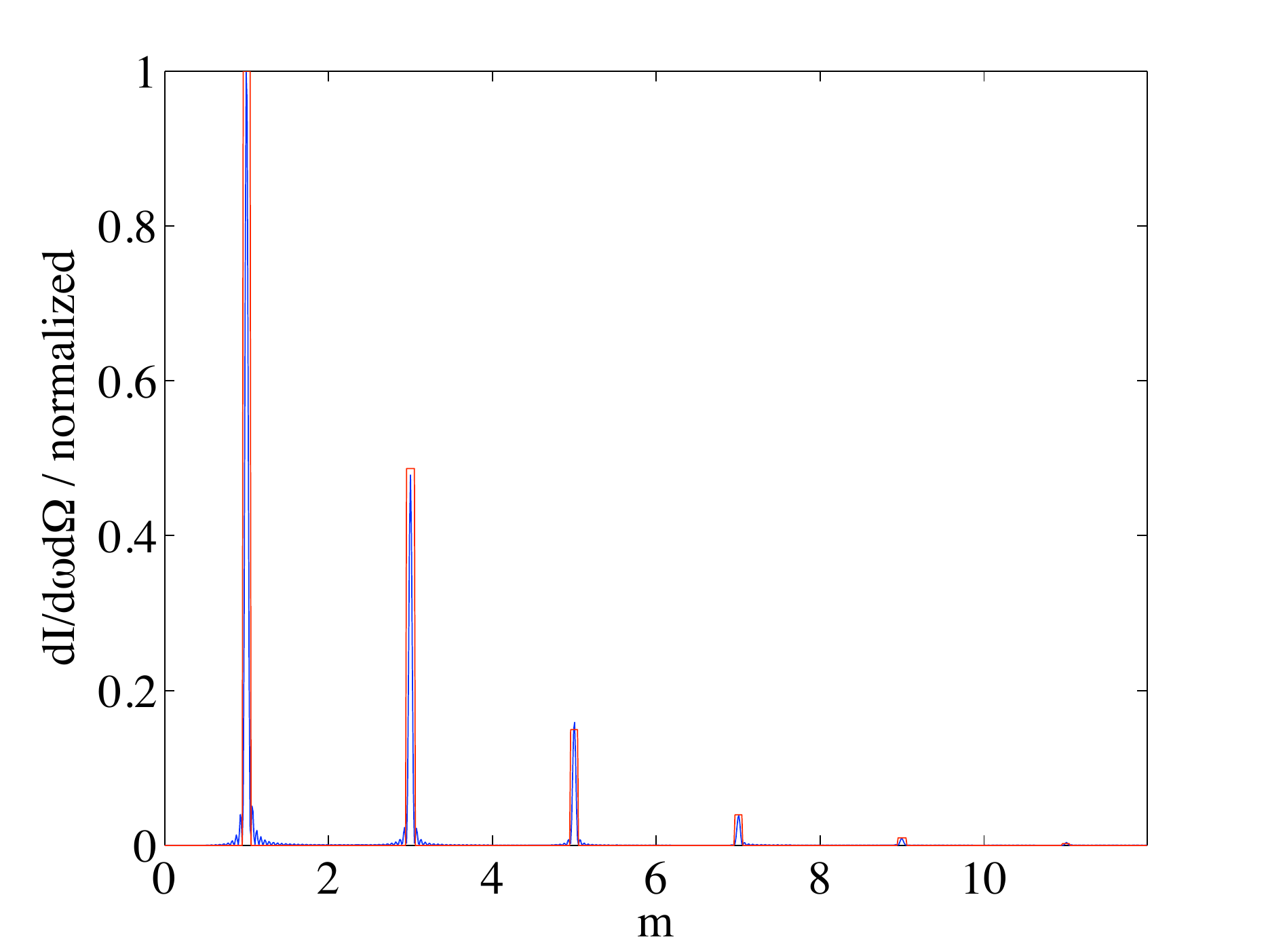}
\caption{The power in harmonics of the fundamental frequency, normalized to the first harmonic, due to non-linear Thomson backscattering from a 5~MeV electron beam colliding with a linear polarized plane wave with a normalized vector potential of $a_0=1$. (red) Calculated from the analytic expression of reference \cite{ISI:000182450200081}. (blue) Calculated from numerical algorithm.}
\label{figure4}
\end{center}
\end{figure}
where $J_\nu(x)$ is a Bessel function of the first kind and $\zeta=a_0^2/(2a_0^2+4)$. In figure \ref{figure4}, the power in harmonics of the fundamental frequency, normalized to the first harmonic, as a function of harmonic $m$, due to non-linear Thomson backscattering from a 5~MeV electron beam colliding with 10 periods of a linearly polarized plane wave with a normalized vector potential of $a_0=1$ is shown. The temporal resolution is such that $\omega_0\Delta\tau=\pi/50$, which represents only 100 sample points to calculate the spectrum from. Also shown is the (normalized) power from the analytic solution. Again the analytic solution models an infinite plane wave solution, and hence the harmonics are discrete. In the numerical solution, a finite number of periods is calculated, and therefore the harmonics have a ${\rm sinc}^2(\omega-\omega_m)$ shape. The scaling of the amplitude of the peaks and the position of the harmonics agrees well between the two solutions. For the numerical solution, the spectrum was calculated as a function of frequency and then divided by the analytically calculated frequency $\omega_1$ to give the horizontal axis. The 12th harmonic corresponds to a frequency of $784\omega_0$, which for an $800~nm$ laser interaction corresponds to a photon energy of 1.2~keV. 
\begin{figure}[htbp]
\begin{center}
\includegraphics[width=4in]{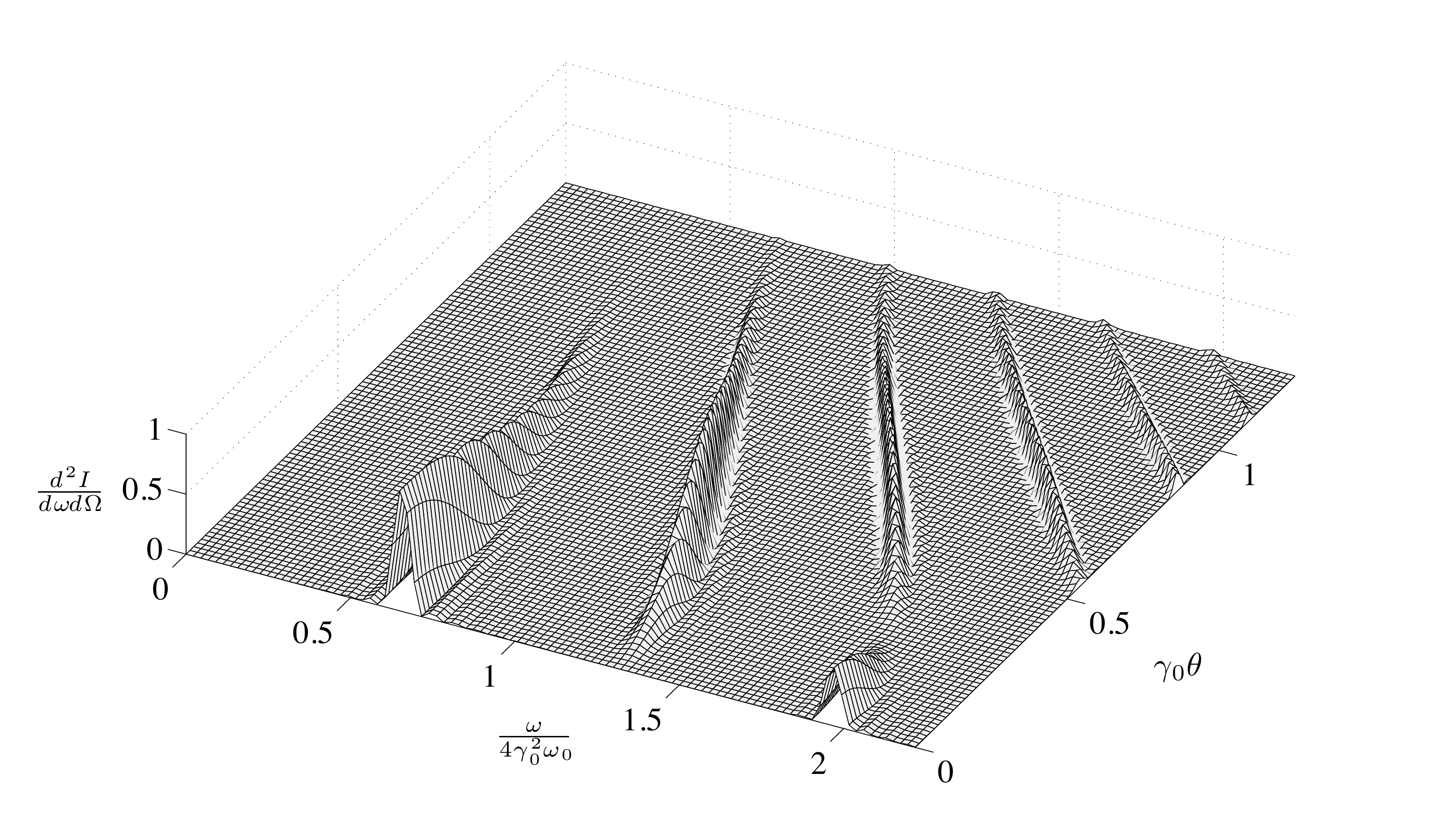}
\caption{The normalized spectral intensity as a function of normalized frequency, $\omega/4\gamma_0^2\omega_0$ and angle $\gamma_0\theta_0$ of radiation scattered by a 5~MeV electron beam colliding with a linear polarized plane wave with a normalized vector potential of $a_0=1$. (c.f. reference \cite{ISI:A1993me59300082})}
\label{figure6}
\end{center}
\end{figure}

Figure \ref{figure6} shows the angularly resolved plot of the same calculations as figure \ref{figure4}, and can be compared with figure 2(b) in reference \cite{ISI:A1993me59300082}, which shows a similar plot calculated from an analytic solution. Note that in their figure, only the first three harmonics were calculated. In the top right of the mesh-plot in figure \ref{figure6}, the next 3 harmonics can be observed. The finite width of the pulse, in both the numerical calculations here and the analytic calculations of reference \cite{ISI:A1993me59300082}, results in the ${\rm sinc}^2(\omega-\omega_m)$ shape of the spectral peaks.

It should be noted that the radiation reaction force model of reference \cite{ISI:000255457000076} has also been included to particle motions for other laser-particle interactions, and the energy loss by the particle compared with the energy in the calculated spectra, with good agreement. The question arises as to what resolution is required to accurately reproduce spectra using this method. Firstly, the particle trajectories themselves have to be reproduced accurately. Secondly, the quadratic interpolation must be an accurate representation of the function $\kappa_\alpha x^\alpha$. Since the calculation is an integral, slight discontinuities at grid points are not as important as the accuracy of the function between grid-points. This means that a spline interpolation method is not necessarily better, and has been found to be worse, than an interpolation that only relates to local grid-points. If the coefficient for a cubic interpolation term is small, $\chi_{3n}\tau^3\ll 1$, then the spectrum should be accurately modeled. The result of this is that the algorithm is not particularly efficient for calculations of highly relativistic particles ($\gg$GeV) performing large radius orbits such as in a classic synchrotron, although it will yield correct results given sufficient resolution (the required resolution for such orbits scales as $\gamma$, because the trigonometric nature of the motion means a cubic term exists and $\kappa_\alpha x^\alpha$ scales with $\gamma$). For such calculations other methods may be preferred. However, for laser or plasma based accelerator interactions, for example, where typically the electron energies are $<10$~GeV but the oscillation frequencies are very fast, $10^{13}-10^{16}$~Hz, the algorithm works well.

\end{document}